# Bell tests with optimal local hidden variable models

Fuming Wang

Department of Physics, Xiamen University, Xiamen, Fujian 361005, China

**Abstract**

An alternative method of detection-loophole-free Bell test is proposed using local hidden variable (LHV) models with optimal detection efficiencies. A framework for constructing such optimal LHV models is presented. Optimal LHV models for maximally and non-maximally entangled twopartite states are constructed to reproduce the quantum correlations within the critical detection efficiencies. The LHV models are shown to be completely equivalent with the existing twopartite Bell inequalities in their optimized setups, and to have even lower critical efficiencies in the LHV modes' own optimized setups. Applications in Bell tests and in device-independent quantum information processing are discussed.

## I. INTRODUCTION

Einstein-Podolsky-Rosen (EPR) experiments display nonlocal correlations between space-like separated measurements [1, 2]. Quantum nonlocality is one of the most peculiar and at the same time one of the most important feature of quantum theory. Bell and other researchers have derived many inequalities that put limits on the correlations producible by local hidden variable (LHV) theories [3-8]. Violation of these limits by the quantum systems would refute local realism and confirm nonlocality in quantum systems. The establishment of a clear boundary between quantum and classical (local realistic) correlations can bring deeper understanding to the quantum mechanics [9-12], and lead to practical applications in the field of quantum information processing, such as device-independent quantum key distribution [13,14] and device-independent randomness generation [15,16].

It is rather difficult to demonstrate quantum nonlocal correlations beyond any possibility of local realistic interpretations. One has to close many loopholes that could lead to possible local realistic interpretations [17-20]. The two most important loopholes are the detection loophole and the locality loophole [20]. The detection loophole is caused by non-ideal detection efficiency in experiments, which have to rely on the "fair sampling" assumption to justify partial instead of full sample collections; the locality loophole refers to the proximity in space-time between the registrations of the correlating signals, leaving the possibility of local communications between the detectors. The technical requirements for removing the detection and the locality loopholes contradict with each other. To remove the detection loophole, one need to place the detectors closer to the source in order to increase the detection efficiency; on the other hand, one need to place the detectors away from the source to remove the locality loophole by creating space-time separated registering events.

Earlier Bell tests have rather low detection efficiencies and can not remove the detection or the locality loophole [21-23]; later on, the locality loophole is remove in some experiments [24,25]. The detection loophole is first removed in an ion-based Bell test [26], and it is removed only recently in photon-based Bell tests [27,28]. The removal of the detection loophole in photon-based Bell tests is very important as they are the most likely kind of Bell tests that can remove both of the detection and the locality loopholes, an achievement that has not been possible so far. Detectors with higher efficiencies and Bell test methodologies with lower critical efficiencies will help achieving this goal.

To carryout detection-loophole-free Bell tests, one needs Bell inequalities that do not assume 100% detection efficiency. Measurement setting dependent threshold efficiencies can be inferred from such inequalities, which give the minimum efficiencies required to distinguish correlation predictions from quantum mechanics and LHV theories. Measurement settings can be optimized to get the minimum threshold efficiency, which is referred to as the critical efficiency of the inequality. Assuming independent detection errors, Garg and Mermin [6] extended the Clauser-Horne-Shimony-Holt (CHSH) Bell inequality to incorporate the effects of non-detection events, and a critical efficiency of 82.8% is obtained for maximally entangled states. Later on, Larsson [8] derived a CHSH-like inequality without the independent error assumption and a slightly higher critical efficiency of 85.4% is found. Clauser and Horne [5] derived a Bell inequality (CH) that incorporated both the joint and the single detection probabilities, and a critical efficiency of 82.8% for the maximally entangled states is found as expected. Later on, Eberhard arrived at essentially the same inequality; it is shown that lower critical efficiencies for the non-maximally entangled states are possible, and the lowest critical efficiency of 66.7% is found for the slightly entangled states [7]. Using a non-maximally entangled state with the CH-Eberhard inequality, one can reach a much lower critical efficiency than using a maximally entangled one. Recently, Giustina *et al.* and Christensen *et al.* reported photon-based detection-loophole-free Bell tests using this method, achieving detection efficiencies around 75% over critical efficiencies around 71% [27-28].

Unlike Bell inequalities which only give local realistic limits to the correlations, LHV models present a direct local realistic realization of the single detection probabilities (SDPs) and the joint detection probabilities (JDPs) in the entangled system. Within certain efficiency limits, some of the existing LHV models have already been able to reproduce the JDPs and the SDPs of an entangled twopartite system [29,30]. We shall refer to such LHV models as proper LHV models.

It is easy to adjust a proper LHV model to have lower detection efficiencies; the opposite is rather difficult. Every proper LHV model has a detection efficiency limit that can not be surpassed without compromising its ability to reproduce the JDPs or the SDPs. This efficiency limit has similar meanings as the critical efficiency of a Bell inequality. We shall refer to this efficiency limit as the critical efficiency of the proper LHV model.

A LHV model may need to reproduce JDPs for arbitrary measurement settings or it may need to only reproduce JDPs for certain limited measurement settings. The difference in the required JDPs will affect the inner workings of the model and its critical efficiency. Therefore, only LHV models that correspond to the same set of required JDPs can be compared meaningfully. The proper LHV model that has the highest critical efficiency among the proper LHV models for the same set of required JDPs is special, because one can refute all proper LHV models by refuting just this particular one LHV model with an efficiency exceeding its critical efficiency. We shall refer to it as an optimal LHV model.

The above discussions present a new method of carrying out Bell tests. Instead of refuting Bell inequalities, one can try to refute local realism by the refuting of optimal LHV models. Since the critical efficiencies of both Bell inequalities and LHV models happen at extreme conditions came out of the same basic local realistic assumptions, one expect agreements between them for the same experimental setup. Conversely, if a LHV model gives the same critical efficiency as a Bell inequality from the same setup, one has to conclude that the LHV model is an optimal LHV model for those JDPs.

It would be less meaningful if the optimal LHV models present only equivalent Bell tests as the existing Bell inequalities; however it will be shown this is not the case. In this paper, we will show that optimal LHV models have identical critical efficiencies as the Bell inequalities at their maximum violation settings, and have even critical efficiencies at the settings optimized for the LHV models.

The paper is structured as the following. A framework for constructing optimal LHV models systematically is presented in Sec. II; Examples of optimal LHV models for maximally entangled states are presented in Sec. III; and that for non-maximally entangled states are presented in Sec. IV. The applications of the presented results are discussed in Sec. V.

## II. CONSTRUCTING OPTIMAL LHV MODELS

Here, we present a framework that can be used to systematically construct optimal LHV models. We use polarization entangled photons as an example of the twopartite systems, and use two identical detectors A and B, each of which has two signal channels label with + and - signs. Independent errors are assumed for the system; the detection efficiency $\eta$ should be identical and invariant for both detectors, and the coincidence efficiency $\eta_{AB}$ satisfies $\eta_{AB} = \eta^2$.

We start from Bell's LHV framework [3] and calculate the JDPs from equation

$$P_{xy}^{LHV}(a,b) = V\left(A_x(a) \cap B_y(b)\right)/V(\Lambda). \qquad (1)$$

Here $x$ and $y$ represent the + and - channels, $a$ and $b$ represent the measurement settings at detectors A and B; $A_x(a)$ and $B_y(b)$ are subsets of the hidden variable sample space $\Lambda$ that would register in the *x* and *y* channels when measured at *a* and *b* settings; function $V(\cdot)$ calculates the volume (area) of a subset in $\Lambda$. For the following discussions, we define subset $A(a) = A_+(a) \cup A_-(a)$ and $B(b) = B_+(b) \cup B_-(b)$ to describe detection events at detector A and B with settings *a* and *b*, and subset $A^c(a)$ and $B^c(b)$ to describe non-detection events at the two settings.

To produce quantum-like JDPs $P_{xy}^{QM}(a,b)$ in an EPR experiment, a portion of the hidden variable samples need to be undetectable [8,29-34]. It means that at least one of $A^c(a)$ or $B^c(b)$ has to be nonempty, and that a proper LHV model can never have 100% detection efficiency for EPR experiments. This is the fundamental difference between LHV and quantum theory, which predicts nonlocal correlations with 100% theoretical efficiency.

The choice of having one or both of $A^c(a)$ and $B^c(b)$ as nonempty can affect the degree of difficulty for the construction of the model. In appendix A, we show that by setting only one of $A^c(a)$ or $B^c(b)$ to nonempty, the constraints that a LHV model has to satisfy to reproduce the required JDPs are reduced by half. Based on this conclusion, we have the following proposition.

*Proposition 1: an optimal LHV model can always be constructed from proper LHV models with an empty $A^c(a)$.*

The proof for *Theorem 1* is quite simple. Since proper LHV models with empty $A^c(a)$ have less number of constraints to satisfy to produce the required JDPs, they will have more degrees of freedom left for efficiency optimization. There is no reason to exclude them from achieving the highest detection efficiency. The ability to construct LHV models equivalent to existing Bell inequalities will confirm this argument.

With and empty $A^c(a)$, we will be able to obtain the shapes of $A_x(a)$ and $B_y(b)$ directly from the required quantum JDPs $P_{xy}^{QM}(a,b)$. We start with a LHV model that needs to simulate $P_{xy}^{QM}(a,b)$ at arbitrary $a$ and $b$ settings. First, we put $P_{xy}^{QM}(a,b)$ in an integral form

$$P_{xy}^{QM}(a,b) = P_{xy}^{QM}(a_0,b) + \int_{a_0}^{a} \frac{\partial P_{xy}^{QM}(a,b)}{\partial a} da, \qquad (2)$$

where $a_0$ is a minimum point for $P_{xy}^{QM}(a,b)$ with fixed $b$. $a_0$ is often also the zero point of $P_{xy}^{QM}(a,b)$, and Eq. (2) now becomes

$$P_{xy}^{QM}(a,b) = \int_{a_0}^{a} \frac{\partial P_{xy}^{QM}(a,b)}{\partial a} da. \qquad (3)$$

When $P_{xy}^{QM}(a_0,b)$ is non-zero, it is often possible to separate $P_{xy}^{QM}(a,b)$ into multiple non-negative terms, and treat them separately as independent detection events (see appendix B). Here, we concentrate on cases suitable for Eq. (3). Next, we map setting $a$ directly to local hidden variable $\lambda$, and carryout the integration through statistical accumulation of joint detection events, and we will have

$$P_{xy}^{LHV}(a,b) = \frac{1}{V(\Lambda)} \int_{\min(a,a_0)}^{\max(a,a_0)} \left| \frac{\partial P_{xy}^{QM}(\lambda,b)}{\partial \lambda} \right| d\lambda \qquad (4)$$

where $P_{xy}^{LHV}(a,b)$ is the JDPs produced from the LHV model and $\left|\partial P_{xy}^{QM}(\lambda,b)/\partial \lambda\right|$ can be view as the probability density function (PDF) of the statistical accumulation process. No extreme point can be present within the integration domain, otherwise one would encounter negative probability during the accumulation process. It should always be possible to locate a minimum point $a_0$ near $a$ without going through other extreme points. Assuming $a_0 < a$, one can obtain the desired $P_{xy}^{LHV}(a,b)$ with the following implementation of subsets

$$A_x(\lambda;a) = \begin{cases} 1, & \lambda \leq a \\ 0, & \lambda > a \end{cases} \qquad (5a)$$

$$B_y(\lambda;b) = \begin{cases} \left|\dfrac{\partial P_{xy}^{QM}(\lambda,b)}{\partial \lambda}\right|, & \lambda \geq a_0(b) \\ 0, & \lambda < a_0(b) \end{cases}. \qquad (5b)$$

Clearly, subset $A_x(a)$ and $B_y(b)$ can be implemented locally since they only depend on local settings and the common hidden variables. Similar implementation can be devised for the case of $a_0 > a$. It is often possible to design a sample space such that the integration of all cases for all the four JDPs can be carried out cooperatively and simultaneously. We will see such examples in Sec. III and IV.

Assuming that a hidden variable sample space $\Lambda$ with subsets $A_x(a)$ and $B_y(b)$ is constructed to reproduce JDPs $P_{xy}^{QM}(a,b)$ through Eq. (4), we will have

$$P_{xy}^{LHV}(a,b) = \eta_B P_{xy}^{QM}(a,b), \quad (6a)$$

$$P_x^{LHV}(a) = P_x^{QM}(a), \quad (6b)$$

$$P_y^{LHV}(b) = \eta_B P_y^{QM}(b), \quad (6c)$$

where $\eta_B = V(B(b))/V(\Lambda)$. We can always set the volume (area) of $B(b)$ to 1, and have a simpler expression of $\eta_B = 1/S$, where $S$ is the volume of the sample space $V(\Lambda)$. We can see that the detection efficiencies of the two detectors are invariant but not identical. Since $B^c(b)$ is nonempty, $V(A(a))$ will always be greater than $V(B(b))$, and the efficiencies can never be identical within sample space $\Lambda$. The only way and the optimal way (following [29,30]) of symmetrizing the efficiencies is to create a mirror image $\Lambda'$ to the sample space $\Lambda$. In the new sample space $\Lambda'$, $A^c(a)$ is nonempty and $A(a)$ and $A^c(a)$ are mirror images of $B(b)$ and $B^c(b)$ in $\Lambda$. Combining the two sample spaces to create a composite sample space and obtain identical efficiencies for both detectors. Eq. (6) now become

$$P_{xy}^{LHV}(a,b) = \eta_{AB} P_{xy}^{QM}(a,b), \quad (7a)$$

$$P_x^{LHV}(a) = \eta_1 P_x^{QM}(a), \quad (7b)$$

$$P_y^{LHV}(b) = \eta_1 P_y^{QM}(b), \quad (7c)$$

where

$$\eta_{AB} = \frac{1}{S}, \quad (8a)$$

$$\eta_1 = \frac{S+1}{2S}. \qquad (8b)$$

We can see that detection efficiencies are invariant and identical in this symmetric LHV model. It is often the case that the efficiencies obtained from such a model does not satisfy the independent error condition $\eta_{AB} = \eta_1^2$. However, it is quite straight forward to derive from it a new model that do satisfy the condition, provided $\eta_{AB}/\eta_1^2 < 1$. Suitable amount of background events not detectable by both detectors ($\lambda \in A^c(a) \cap B^c(b)$) can be added to the sample space. The relative amplitudes among the JDPs will not be affected but their absolute amplitudes will be reduced. The new detection and coincidence efficiencies would become $c\eta_1$ and $c\eta_{AB}$, where $c < 1$ is a reduction factor. Solving $c^2\eta_1^2 = c\eta_{AB}$, a new detection efficiency

$$\eta_2 = \frac{2}{S+1}, \qquad (9)$$

can be obtained. Replacing $\eta_1$ and $\eta_{AB}$ in Eq. (7) with $\eta_2$ and $\eta_2^2$, we will have an optimal LHV model that has independent errors.

Another type of efficiency, the conditional detection efficiency [8], can be defined as $P(A(a)|B(b))$ or $P(B(a)|A(b))$, which are identical for our models. For symmetric models without the assumption of independent errors (Eq. (7)), the conditional detection efficiency can be calculated with Eq. (9).

We now try to prove a proposition that can be used to test whether a proper LHV model is optimal.

*Proposition 2: a proper LHV model is optimal if it has an $A_x(a)$ that is a proper subset of $E = \bigcup_b B(b)$.*

Here $A_x(a)$ and $B(b)$ are subsets in the original samples space $\Lambda$ before it is symmetrized, and $E$ is the envelope that encloses all $B(b)$ in $\Lambda$. The proof is simple. We know that $A(a)$ has to enclose $E$ to reproduce the required JDPs, i.e. $A(a) \supseteq E$. Therefore, only subset $A(a) \cap E^c$ can be optimized for detection efficiency. If there is an $A_x(a) \subseteq E$, then all samples in $A(a) \cap E^c$ belong to subset $A_{x'}(a)$ for the opposite channel of detector A. Since the ratio between $V(A_x(a))$ and $V(A_{x'}(a))$ must equal to $P_x^{QM}(a)/P_{x'}^{QM}(a)$, $V(A(a) \cap E^c)$ is fixed also and can not be further optimized. Therefore, the proper LHV model must be optimal.

### III. LHV MODELS FOR MAXIMALLY ENTANGLED STATES

We now try to construct optimal LHV models for maximally entangled photons $|\Psi\rangle = (|H\rangle_A |H\rangle_B + |V\rangle_A |V\rangle_B)/\sqrt{2}$, where $|H\rangle$ and $|V\rangle$ are the horizontal and the vertical polarization states. Several LHV models for similar systems have been reported before [29-34]. Here, we would like to demonstrate that essentially the same LHV model can be constructed systematically with our framework, and that the constructed LHV models can be shown to be optimal. First, we try to construct an optimal LHV model for arbitrary $a$ and $b$ settings. We shall refer to such LHV models as NxN models, since they require N measurements at each detector to verify the correlations. N could be a fairly large number if a high statistical significance is desired. We start from quantum mechanics JDP

$$P_{++}^{QM}(a,b) = \frac{1}{2}\cos^2(a-b),$$

and construct the PDF following the framework as

$$\left|\frac{\partial P_{++}^{QM}(\lambda,b)}{\partial \lambda}\right| = \frac{1}{2}|\sin 2(\lambda-b)|. \tag{10}$$

The relevant minimum points (also zero points) of $P_{++}^{QM}(a,b)$ are at $b \pm \pi/2$, and there also is a maximum point at $b$. For $a > b$, the closest minimum point is $b + \pi/2$ and should be selected as the upper bound of the integration. According to Eq. (4)

$$P_{++}^{LHV}(a,b) = \frac{1}{V(\Lambda)} \int_a^{b+\frac{\pi}{2}} \frac{1}{2}|\sin 2(\lambda-b)| d\lambda. \tag{11}$$

For $a < b$, the closest minimum point is $b - \pi/2$, and the integration becomes

$$P_{++}^{LHV}(a,b) = \frac{1}{V(\Lambda)} \int_{b-\frac{\pi}{2}}^a \frac{1}{2}|\sin 2(\lambda-b)| d\lambda. \tag{12}$$

Map hidden variable $\lambda$ in Eq. (12) to $\lambda + \pi/2$ and we obtain an identical integration

$$P_{++}^{LHV}(a,b) = \frac{1}{V(\Lambda)} \int_{b}^{a+\frac{\pi}{2}} \frac{1}{2} |\sin 2(\lambda - b)| d\lambda. \qquad (13)$$

With this change, we can construct subset $A_+(a)$ and $B_+(b)$ for both Eq. (11) and Eq. (13) as

$$A_+(\lambda;a) = \begin{cases} 1, & a \leq \lambda \leq a + \frac{\pi}{2} \\ 0, & otherwise \end{cases}, \qquad (14a)$$

$$B_+(\lambda;b) = \begin{cases} \frac{1}{2}|\sin 2(\lambda - b)|, & b \leq \lambda \leq b + \frac{\pi}{2} \\ 0, & otherwise \end{cases}. \qquad (14b)$$

Subset $A_-(a)$ and $B_-(b)$ for $P_{--}^{LHV}(a,b)$ can be implemented similarly, and so do the other two JDPs. It is possible to design a sample space with subsets $A_x(a)$ and $B_y(b)$ that would integrate for all four JDPs simultaneously. Fig. 1 shows such a sample space, with the green (+) and orange (-) sinusoids correspond to subset $B_+(b)$ and $B_-(b)$, the two parallel boxes represent subset $A_+(a)$ and $A_-(a)$, and the gray background labeled with $\phi$ represent subset $B^c(b)$. The area of the whole sample space $S$ equal to $\pi/2$. A symmetric LHV model with independent errors can be constructed from the asymmetric model in Fig. 1 as discussed in Sec. II, and its detection efficiency is 77.8%, calculated with Eq. (9).

Since the sinusoids representing $B_+(b)$ and $B_-(b)$ in Fig. 1 move continuously through the sample space, the envelope of $B(b)$ is the whole sample space; therefore every subset $A_+(a)$ is a proper subset of the envelope, and Fig. 1 is an optimal LHV model according to *proposition 2*. In Fig. 1 and in models that will be discussed, the shapes of subset $B_+(b)$ and $B_-(b)$ are determined completely by the four JDPs and by the boundary between subset $A_+(a)$ and $A_-(a)$. A different boundary between $A_+(b)$ and $A_-(b)$ would produce shear deformations to $B_+(b)$ and $B_-(b)$, but the volumes (areas) of the subsets should be identical due to Cavalieri's principle [38]. For the purpose of efficiency calculation, one can consider subset $B_+(b)$ and $B_-(b)$ completely determined by the four JDPs.

To compare with Bell inequalities, we have to restrict the measurement settings to 2x2 measurements (two measurements at each detector). For the maximum violation of CHSH type inequalities and the minimum threshold efficiencies, one chooses setting *a* and *b* that satisfy $|a-b|=\pi/8$. Since the model only need to produce JDPs at these limited settings, the sinusoidal shapes of $B_+(b)$ and $B_-(b)$ are no longer required. To further reduce the area of the sample space in order to maximize the detection efficiency, we change $B_+(b)$ and $B_-(b)$ into polygons as shown in Fig. 2. The widths and the heights of the rectangles within the polygons are completely determined by the choice of *a* and *b*, and by the corresponding JDPs. The width of the two areas labeled by $\phi$ is $\pi/4$, and the shifting of $B_+(b)$ and $B_-(b)$ to the left and the right by $\pi/8$ happens to over them, making the the envelope of $B(b)$ equal to the whole sample space. Therefore, according to *proposition 2,* Fig. 2 is also an optimal LHV model.

The area of the sample space in Fig. 2 is $\sqrt{2}$, and the independent detection efficiency of the model is 82.8%, which is exactly the same as the critical efficiencies of Garg and Mermin's version of CHSH and the CH-Eberhard inequalities. This demonstrates the equivalence between the optimal model in Fig. 2 and the CHSH and the CH-Eberhard Bell inequalities. This equivalence can be further proved by comparing with Larsson's results which do not require the assumption of independent errors. The detection efficiency for Fig. 2 without the requirement of independent errors is calculated with Eq. (8) to be 85.4%, which matches the critical efficiency of Larsson's version of the CHSH inequality [16]. Furthermore, in the same paper, Larsson derived another CHSH-like inequality for conditional detection efficiency, which has a critical value of 82.8%. The same type of efficiency for the LHV model can be calculated with Eq. (9) and the same value of 82.8% is obtained as expected. The agreement of the critical efficiencies with the Bell inequalities suggests that the LHV model in Fig. 2 is an optimal LHV model.

In conclusion, the LHV models in Figs. 1 and 2 are optimal LHV models; the model in Fig. 2 has the same critical efficiency as the existing Bell inequalities for maximally entangled twopartite systems.

### IV. LHV MODELS FOR NON-MAXIMALLY ENTANGLED STATES

#### A. NxN model

We now try to construct optimal LHV models for non-maximally entangled photons $|\Psi\rangle = (r|H\rangle_A|H\rangle_B + |V\rangle_A|V\rangle_B)/\sqrt{1+r^2}$, where parameter $r \in (0,1]$ controls the degree of the entanglement. When $r=1$, the state reverts back to a maximally entangled one. Now we try to construct a NxN LHV model. We start from quantum mechanics JDP

$$P_{++}^{QM}(a,b) = (r^2 \cos^2 a \cos^2 b + \sin^2 a \sin^2 b + 0.5r \sin 2a \sin 2b)/(1+r^2)$$

and obtain the PDF as

$$\left| \frac{\partial P_{++}^{QM}(\lambda,b)}{\partial \lambda} \right| = P_+(b)|\sin 2(\lambda - \theta_+(b))| \tag{15}$$

where

$$P_+(b) = \frac{r^2 \cos^2 b + \sin^2 b}{1+r^2} \tag{16}$$

is the single detection probability at the '+' channel of detector B, and $\theta_+(b)$ is defined as

$$\cos 2\theta_+(b) = \frac{r^2 \cos^2 b - \sin^2 b}{r^2 \cos^2 b + \sin^2 b} \tag{17a}$$

$$\sin 2\theta_+(b) = \frac{2r \sin b \cos b}{r^2 \cos^2 b + \sin^2 b} \tag{17b}$$

For parameter r = 1, $P_+(b) = 0.5$, $\theta_+(b) = b$, and we return to Eq. (10). The corresponding minimum points (also zero points) of $P_{++}^{QM}(a,b)$ are at $\theta_+(b) \pm \pi/2$, and there also is a maximum point at $\theta_+(b)$. Similar to Eq. (14), subset $A_+(a)$ and $B_+(b)$ for $P_{++}^{LHV}(a,b)$

can be defined as

$$A_+(\lambda;a) = \begin{cases} 1, & a \leq \lambda \leq a + \dfrac{\pi}{2} \\ 0, & otherwise \end{cases} \quad (18a)$$

$$B_+(\lambda;b) = \begin{cases} P_+(b)|\sin 2(\lambda - \theta_+(b))|, & \theta_+ \leq \lambda \leq \theta_+ + \dfrac{\pi}{2} \\ 0, & otherwise \end{cases} \quad (18b)$$

The corresponding subset $A_-(a)$ and $B_-(b)$ for $P_{--}^{LHV}(a,b)$ are different in both amplitude and phase

$$A_-(\lambda;a) = \begin{cases} 1, & a + \dfrac{\pi}{2} \leq \lambda \leq a + \pi \\ 0, & otherwise \end{cases} \quad (19a)$$

$$B_-(\lambda;b) = \begin{cases} P_-(b)|\sin 2(\lambda - \theta_-(b))|, & \theta_- + \dfrac{\pi}{2} \leq \lambda \leq \theta_- + \pi \\ 0, & otherwise \end{cases} \quad (19b)$$

where

$$P_-(b) = \frac{r^2 \sin^2 b + \cos^2 b}{1 + r^2} \quad (20)$$

is the single detection probability at the '-' channel of detector B, and $\theta_+(b)$ is defined as

$$\cos 2\theta_-(b) = \frac{-r^2 \sin^2 b + \cos^2 b}{r^2 \sin^2 b + \cos^2 b} \quad (21a)$$

$$\sin 2\theta_-(b) = \frac{2r \sin b \cos b}{r^2 \sin^2 b + \cos^2 b} \quad (21b)$$

We have also shifted $A_-(a)$ and $B_-(b)$ for $\pi/2$ to place them along $A_+(a)$ and $B_+(b)$.

Instead of a rectangle, the envelope of $B(b)$ in this model has a sinusoidal shape. A sample space is constructed from it and is shown in Fig. 3. The green (+) and orange (-) sinusoids correspond to subset $B_+(b)$ and $B_-(b)$, and the gray background labeled with $\phi_1$ and $\phi_2$ represents subset $B^c(b)$. The envelope covers the green (+) sinusoid, the orange (-) sinusoid, and the $\phi_1$ dark gray background. $\phi_2$ is needed to produce the correct distribution for $P_x^{LHV}(a)$, and its shape is determined by $P_x^{QM}(a)$. We can see that the $A_+(a)$ shown in Fig. 3 is enclosed completely in the envelope hence this model is an optimal LHV model as required by *Proposition 2*. No additional $\phi_2$ background is needed for LHV models of maximally entangled states (Figs. 1 and 2) because their $P_+^{QM}(a)$ and $P_-^{QM}(a)$ always equal to 0.5.

Assuming the area of the envelope between 0 and $\pi/2$ is $S_1$, the total area of the sample space $S$ would equal to $S_1(1+r^2)/r^2$. A computer program is developed to calculate $S_1$ numerically. The detection and coincidence efficiencies are calculated from $S_1$ and are plotted as functions of $r$ in Fig. 4. As $r$ approaching zero, the critical efficiencies approach to a limiting value of zero. For comparison, the CH-Eberhard inequality approaches a limiting critical efficiency of 0.667 as $r$ approaching zero.

### B. 2x2 and 3x3 models

To compare with the CH-Eberhard inequality directly, we need to construct an optimal LHV model that makes 2x2 measurements instead of NxN measurements. The procedure for constructing such a model is similar. Instead of constructing an envelope of $B(b)$ with arbitrary setting $b$, we can build an envelope out of two settings $b_1$ and $b_2$. The shape and the area $S$ of sample space can be determined by requiring the correct distribution of $P_x^{LHV}(a)$ for settings $a_1$ and $a_2$. Since the 2x2 models only need to produce JDPs at four possible measurement settings, the sample space can be constructed as polygons instead of sinusoids without any effects on the JDPs. Fig. 5 shows a LHV model with $a_i = 0, \pi/4$ and $b_j = \pm\pi/16$, where the smooth envelope in Fig. 3 is replaced by polygons (area other than the $\phi_2$ background). The displayed $A_+(a)$ in Fig. 5 is completely within the envelope hence Fig. 5 is an optimal LHV model according to *Proposition 2*.

Critical efficiencies for the CH-Eberhard inequality $\eta^{CH}(r)$ and that for the LHV model

$\eta^{LHV}(r)$ are searched exhaustively for a series of *r*'s with a step size of $\pi/200$ in settings $a_i$ and $b_j$. It is found within the search accuracy that $\eta^{CH}(r)$ can always be reached by letting $a_1 = -b_1$, and $\eta^{LHV}(r)$ by letting $a_j = 0, \pi/4$ and $b_1 = -b_2$. More refined searches are carried out with these constraints to reduce the workload. The obtained $\eta^{CH}(r)$ and $\eta^{LHV}(r)$ are plotted in Fig. 6. We see that $\eta^{CH}(r)$ (triangles) and $\eta^{LHV}(r)$ (circles) converge to 0.828 and 0.667 when *r* approaches 1 and 0, while $\eta^{LHV}(r)$ has slightly lower values (around 1%) than $\eta^{CH}(r)$ for the intermediate *r*'s. The inset in Fig. 6 shows a smooth variation of the differences.

In general, $\eta^{CH}(r)$ and $\eta^{LHV}(r)$ are generated by different angle settings. It is found however that an angle setting $\{a_i, b_j\}_{CH}$ that lead to $\eta^{CH}(r)$ generates the same efficiency for the 2x2 LHV model, i.e.

$$\eta^{CH}(r, \{a_i, b_j\}_{CH}) = \eta^{LHV}(r, \{a_i, b_j\}_{CH}).$$

This effect shows that the Bell inequalities are equivalent with the optimal LHV models only at the settings that give critical efficiency or maximum violation to the inequalities. This is exactly the case for the maximally entangled states as has already been observed. The agreement with the CH-Eberhard inequality at its maximum violation conditions confirms that the 2x2 LHV model is indeed an optimal LHV model; at the same time, it shows that the CH-Eberhard inequality is only capable of a constrained optimization of the critical efficiency while the 2x2 LHV model can perform a full optimization and obtain a even lower critical efficiency.

New constraints can be added to Fig. 5, and even lower efficiency can be obtained. We increase measurements at each device from 2 to 3, and construct a 3x3 LHV model similarly as the 2x2 model. Exhaustive searches are run with a step size of $\pi/32$, and critical efficiencies are found when $a_i$ and $b_j$ both belong to $\{0, \pi/8, 3\pi/8\}$. The efficiencies of the 2x2, the 3x3, and the NxN LHV models are plotted in Fig. 7, and significant efficiency reduction from the 2x2 model (triangles) to the 3x3 model (circles) can be observed. Using the testing configuration used by Christensen *et al.* [28], a critical efficiency of 0.708 is obtained from the CH-Eberhard inequality, while a critical efficiency of 0.616 is obtained from the 3x3 LHV model. A reduction of over 9 percent point is achieved by adding just one more measurement at each detector. Further reduction is still possible as there is plenty room between $\eta_{\min}^{LHV}(r)$ from the 3x3 model (circles) and the NxN model (squares).

## V. DISCUSSIONS

We have presented a systematic way of constructing LHV models for arbitrary twopartite entangled systems together with rigorous arguments to show that such constructed LHV models are optimal. The equivalence between the constructed LHV models and the existing Bell inequalities for twopartite systems confirms the methodology and the arguments. It is quite remarkable that two completely different types of realizations of the same local realistic assumptions can lead to completely identical predictions on the critical efficiencies that separate local from non-local correlations. It demonstrates that these boundaries are indeed the intrinsic properties of the local realistic assumptions instead of the peculiarities of inequality proof or model construction.

In Fig. 3, we have, for the first time, a LHV model of a twopartite quantum state with arbitrary amount of entanglement. Fig. 4 gives the lower bound of critical efficiencies for any twopartite Bell tests. With the 2x2 and the 3x3 models, we have, for the first time, testable LHV models that have critical efficiencies lower than those of the Bell inequalities. The reduction in the critical efficiency may help the effort to carryout Bell tests that are free from both the detection and the locality loopholes. Comparing with the inequality based Bell tests, the optimal LHV based Bell tests are less rigorous but more versatile and more informative.

To refute local realism with an optimal LHV model, one needs to show that the measured JDPs and SDPs agree statistically with the quantum theory (*hypothesis I*), We assume that the following quantities, including efficiency $\eta$, the total number of events $n$, the variance for the counts of all types of events σ (assuming they are identical) can be measured, estimated based on the experimental setup or the collected data. The statistical significance of *hypothesis I* can be tested with the following $\chi^2$ values for a NxN model:

$$\chi_1^2(m_1) = \sum_{xy} \sum_{i,j=1}^{N} \left( \frac{n_{ij}^{xy} - \eta^2 n P_{xy}^{QM}(a_i, b_j)}{\sigma} \right)^2 \tag{22a}$$

$$\chi_2^2(m_2) = \sum_{x} \sum_{i=1}^{N} \left( \frac{n_i^x - \eta n P_x^{QM}(a_i)}{\sigma} \right)^2 + \sum_{y} \sum_{j=1}^{N} \left( \frac{n_j^y - \eta n P_y^{QM}(b_j)}{\sigma} \right)^2 \tag{22b}$$

where $m_1 = 4N^2 - 1$ and $m_2 = 4N - 1$ are the degrees of freedom for $\chi_1^2$ and $\chi_2^2$; $n_{ij}^{xy}$, $n_i^x$, $n_j^y$ are the counts from channel $x$ and $y$ at setting $a_i$ and $b_j$; and $P_{xy}^{QM}(a_i, b_j)$, $P_x^{QM}(a_i)$, and $P_y^{QM}(b_j)$ are the joint and single detection probabilities from the quantum theory for the same settings.

Bell tests, especially detection-loophole-free Bell tests, are not only important for deciphering quantum nonlocality, but also for practical applications. Bell inequality violations have been used to design device-independent quantum key distribution protocols [13, 14] and to certify quantum random number generators so that true randomness is verified instead of assumed [15,16]. The existence of Bell inequality violations exclude the possibility of the data being generated from any classical mechanisms, either due to malfunction of the quantum generator or due to tempering from adversaries. Optimal LHV models can be used in such applications also. Verification of the single and joint detection probabilities at efficiency beyond the critical efficiency of an optimal LHV model would exclude any possibility of non-quantum origin for the obtained random bits. With a lower critical efficiency, the LHV model based certification method can make the application more practical.

The frame work can be used to construct LHV models for twopartite entangled systems other than the EPR systems, and test their correlations against local realistic limits. In a delayed-choice experiment proposed Ionicioiu and Terno [36], and carried out by Kaizer *et al.* [37], one of entangled photon is sent through a Mach-Zehnder interferometer (MCI) for self-interfering, while the other one is send through a phase modulator that can rotate its polarization. The goal of the experiment is to exercise control over the self-interfering behavior at the MCI from the phase modulator at a distance and at a delayed time. The detection events are separated in space-time so that no causal connections can be made between them. Yet, correlations are predicted and confirmed. This is completely against the concept of classical locality. However, we are able to quite surprisingly construct an optimal LHV model with 100% efficiency for the system (see Appendix. B). It means that correlations predicted by quantum mechanics and by the LHV theories are indistinguishable. It shows that LHV models can bring new insights to the study of entangled systems.

Overall, the optimal LHV models present a versatile and powerful tool to study or certify nonlocal correlations in entangled systems.

## VI. CONCLUSIONS

A framework for constructing such LHV models is presented together with rigorous arguments to show that the constructed LHV models are optimal. The equivalence between the optimal LHV models and the existing Bell inequalities for twopartite systems confirms the methodology and the arguments. LHV models for non-maximally entangled twopartite states are constructed for the first time and are shown to have optimal efficiencies. Lower bounds of critical efficiencies are established for any twopartite Bell tests. With the 2x2 and the 3x3 models, we have, for the first time, testable LHV testing methods that have critical efficiencies lower than the Bell inequalities. Applications of optimal LHV models in Bell tests and in device-independent quantum information processing are discussed.

## ACKNOWLEDGEMENTS

The Author acknowledges support by the National Basic Research Program of China under grant No. 2012CB933502 and No. 2012CB933503.

## APPENDIX A

We show that a LHV model can reproduce the required JDPs with much less constraints if either $A^c(a)$ or $B^c(b)$ is empty. Here, $A^c(a)$ and $B(b)$ are the non-detection subset for detector A and B in the original samples space $\Lambda$ before it is combined into a symmetric sample space.

To construct a proper LHV model, one has to construct a hidden variable sample space $\Lambda$ with subsets $A_x(a)$ and $B_y(b)$ that can produce the desired JDPs. This task can be accomplished by finding a solution to a system of equations. Let $i$ indexes hidden variable $\lambda$, the 1's of binary variables $B_+(i,b)$, $B_-(i,b)$, and $B_0(i,b)$ represent detection at the '+' channel, detection at the '-' channel, and non-detection at both channels of detector B, and the 0's represent other situations. We have

$$B_x(a) = \{ i \mid \lambda(i) \in \Lambda, B_x(i,b) = 1, \ x \in \{'+','-'\}\} \tag{A1}$$

$$B^c(a) = \{ i \mid \lambda(i) \in \Lambda, B_0(i,b) = 1\} \tag{A2}$$

$$B_+(i,b) + B_-(i,b) + B_0(i,b) = 1, \ i = 1,\cdots,N, \tag{A3}$$

where N is the number of hidden variable samples in sample space $\Lambda$. Eq. (A3) means that $B_+(i,b)$, $B_-(i,b)$, and $B_0(i,b)$ have mutual exclusivity among themselves. Similar binary variables $A_+(i,a)$, $A_-(i,a)$, and $A_0(i,a)$ can be defined for detector A. Assuming that $A_x(i,a)$ are already known, we want to solve $B_y(i,b)$ with a system of linear equations. Before doing that, we need to discretize settings $a$ and $b$ properly into discrete $a_j$ and $b_k$ in order to obtain discrete and independent $A_x(i,a_j)$ and $B_y(i;b_k)$ variables. From Eq. (1) and (7), we have

$$\frac{1}{N}\sum_i A_x(i,a)B_y(i,b) = \eta^2 P_{xy}^{QM}(a,b),$$

where $\eta$ is the detection efficiency of the system. The smallest meaningful change in $a$ or $b$ need to cause the binary variables $A_x(i,a)$ or $B_y(i,b)$ to change its value at least at one hidden variable, which would result in a change of $1/N$ in $\eta^2 P_{xy}^{QM}(a,b)$. The maximum range of $\eta^2 P_{xy}^{LHV}(a,b)$ is approximately 0.0 to 0.4, which would need a maximum number of 0.4N discrete $a$ and $b$ values to cover. The total number of independent variables $B_+(i,b_j)$ and $B_-(i,b_k)$ will be less or equal to $2N \times 0.4N = 0.8N^2$. Variable $B_0(i,b)$ depend on $B_+(i,b_j)$ and $B_-(i,b_k)$ through Eq. (A3). This means that no more than $0.8N^2$ linear equations can be associated with $B_+(i,b_k)$ and $B_-(i,b_k)$, otherwise no solution can be found. This estimation for the number of independent variables is rather crude however we will show that the conclusion drawn does not depend on the exact number of independent variables.

Now, to produce the four desired JDPs, the linear equations that independent variables $B_+(i,b_k)$ and $B_-(i,b_k)$ have to obey the following four sets of linear equations

$$\frac{1}{N}\sum_i A_x(i,a_j)B_y(i,b_k) = \eta^2 P_{xy}^{QM}(a_j,b_k); \quad j,k=1,2,\cdots 0.4N, \quad (A4)$$

where $x,y \in \{'+','-'\}$. The total number of linear equations is $4\times(0.4N)^2 \approx 0.6N^2$. However, if $A^c(a)$ is empty, we can sum over index $x$ on both sides of Eq. (A4) and replace the summation of $A_x(i,a_j)$ with 1. We will have the following two sets of linear equations:

$$\frac{1}{N}\sum_i B_y(i,b_k) = \eta^2 P_y^{QM}(b_k); \quad j,k=1,2,\cdots 0.4N. \quad (A5)$$

Because the Eq. (A5) does not depend on index $j$, the total number of independent linear equations in Eq. (A5) is $0.8N$ instead of $0.3N^2$. Therefore, the total number of independent linear equations in Eq. (A4) becomes $0.32N^2+0.8N \approx 0.3N^2$ instead of $0.6N^2$. The number of constraints is reduced by half. This conclusion does not reply on an accurate estimation of the exact number of constraints. With the significant reduction of the constraints, a LHV model has more degrees of freedom for efficiency optimization and has a better chance of becoming a optimal LHV model.

If $A^c(a)$ is not empty but $A^c(a) \cap B(b)$ is empty, Eq. (A5) is still valid as wherever the summation of $A_x(i,a_j)$ does not equal to 1, $B_y(i,b_k)$ equal to 0. One often adds undetectable background hidden variable samples $\lambda \in A^c(a) \cap B^c(b)$ to the sample space to adjust the detection efficiency while keeping the relative amplitudes of the JDPs unchanged.

# APPENDIX B

In this appendix, we construct an optimal LHV model for a delayed-choice experiment proposed by Ionicioiu and Terno [36], and carried out by Kaizer *et al*. [37]. In the experiment, one of the entangled photons is sent through a MCI for self-interfering, while the other one is send through a phase modulator which can rotate its the polarization. The goal of experiment is to exercise control over the self-interfering behavior at the MCI from the phase modulator at a distance and at a delayed time. The quantum mechanics JDPs for polarization measurements at detectors after the MZI and the modulator are

$$P_{++}^{QM}(a,b) = \frac{1}{4}\cos^2 a + \frac{1}{2}\cos^2 \frac{b}{2}\sin^2 a, \tag{B1a}$$

$$P_{+-}^{QM}(a,b) = \frac{1}{4}\cos^2 a + \frac{1}{2}\sin^2 \frac{b}{2}\sin^2 a, \tag{B1b}$$

$$P_{-+}^{QM}(a,b) = \frac{1}{4}\sin^2 a + \frac{1}{2}\cos^2 \frac{b}{2}\cos^2 a, \tag{B1c}$$

$$P_{--}^{QM}(a,b) = \frac{1}{4}\sin^2 a + \frac{1}{2}\sin^2 \frac{b}{2}\cos^2 a. \tag{B1d}$$

Here $a$ is the angle of polarization rotation after the phase modulator, and $b$ is the phase change introduced by the MZI. The first terms and the second terms on the left-hand side of Eq. (B1) can be grouped into two separate JDPs:

$$P_{++}^{1st}(a,b) = \frac{1}{4}\cos^2 a, \tag{B2a}$$

$$P_{+-}^{1st}(a,b) = \frac{1}{4}\cos^2 a, \tag{B2b}$$

$$P_{-+}^{1st}(a,b) = \frac{1}{4}\sin^2 a, \tag{B2c}$$

$$P_{--}^{1st}(a,b) = \frac{1}{4}\sin^2 a. \tag{B2d}$$

and

$$P_{++}^{2nd}(a,b) = \frac{1}{2}\cos^2 \frac{b}{2}\sin^2 a, \tag{B3a}$$

$$P_{+-}^{2nd}(a,b) = \frac{1}{2}\sin^2 \frac{b}{2}\sin^2 a, \tag{B3b}$$

$$P_{-+}^{2nd}(a,b) = \frac{1}{2}\cos^2 \frac{b}{2}\cos^2 a, \tag{B3c}$$

$$P_{--}^{2nd}(a,b) = \frac{1}{2}\sin^2 \frac{b}{2}\cos^2 a. \tag{B3d}$$

Optimal LHV models for $P_{xy}^{1st}(a,b)$ and $P_{xy}^{2nd}(a,b)$ with arbitrary $a$ and $b$ can be constructed using Eq. (4). The PDFs for $P_{xy}^{1st}(a,b)$ and $P_{xy}^{2nd}(a,b)$ are all proportional to $|\sin 2\lambda|$ and therefore their minimum and zero points are either $0$ or $\pi/2$. Since the minimum points are fixed, the profile of $B(b)$ is also fixed in the samples space and the envelop of $B(b)$ equals to $B(b)$, which can be selected as the whole sample space. Sample space for $P_{xy}^{1st}(a,b)$ and $P_{xy}^{2nd}(a,b)$ are shown in Fig. 8, where the two sinusoids correspond to the two independent JDPs and the green (+) and the orange (-) areas correspond to registrations in the + and – channels of detector B. Since both $A^c(a)$ and $B^c(b)$ are empty in this model, the detection efficiency for both detector A and B are 100%, so is the coincidence efficiency. In other word, the correlations predicted by quantum mechanics and by the LHV theories are indistinguishable in this system. It shows that LHV models can bring new insights to the study of the entangled systems.

---

**FIGURES**

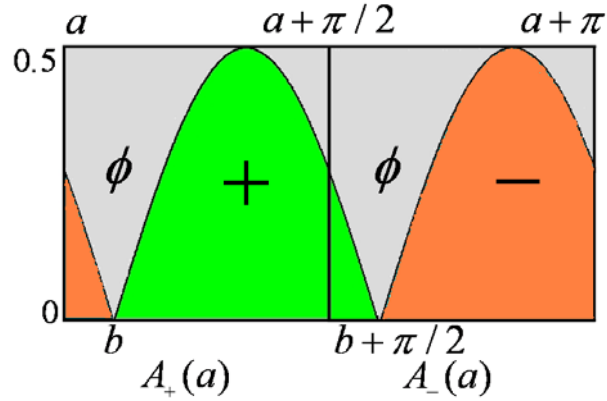

**FIG. 1.** NxN LHV model for maximally entangled quantum states. The two boxes represent subset $A_+(a)$ and $A_-(a)$, the green (+) and orange (-) colored sinusoids represent subset $B_+(b)$ and $B_-(b)$, the grey background ($\phi$) represents subset $B^c(b)$.

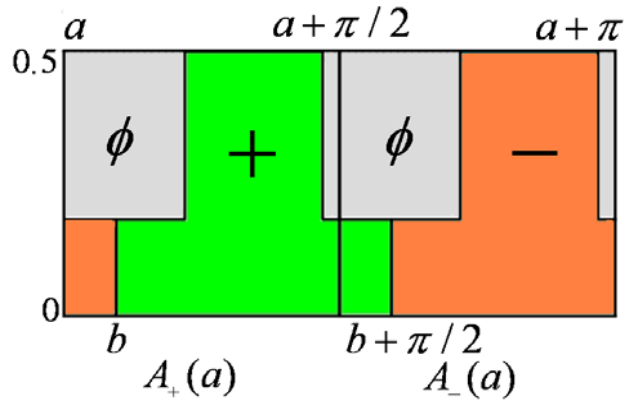

**FIG. 2.** 2x2 LHV model for maximally entangled quantum states. $a$ and $b$ are limited to configurations that satisfy $|a-b| = \pi/8$. The two boxes represent subset $A_+(a)$ and $A_-(a)$, the green (+) and orange (-) colored polygons represent subset $B_+(b)$ and $B_-(b)$, and the grey background (labeled with $\phi$) represents subset $B^c(b)$.

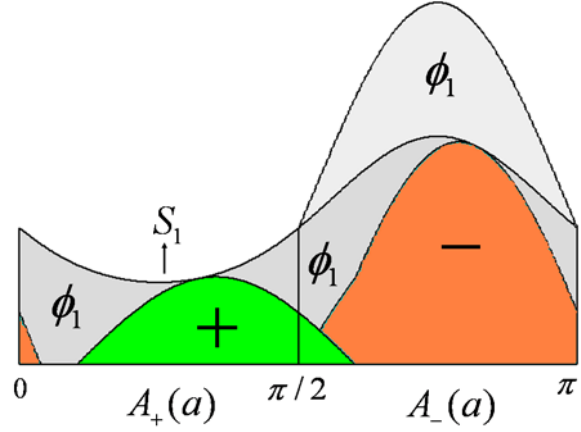

**FIG. 3.** NxN LHV model for non-maximally entangled quantum states. The green (+) and orange (-) colored sinusoids represent subset $B_+(b)$ and $B_-(b)$, and the grey background ($\phi_1$ and $\phi_2$) represents subset $B^c(b)$. The model shown here has $r = 0.60$.

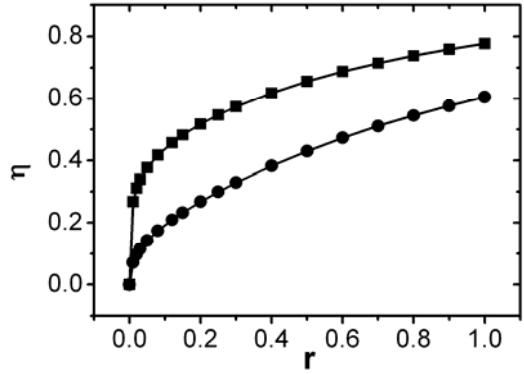

**FIG. 4.** Detection efficiency $\eta$ (squares) and coincidence efficiency $\eta_{AB}$ (circles) as functions of entanglement parameter $r$.

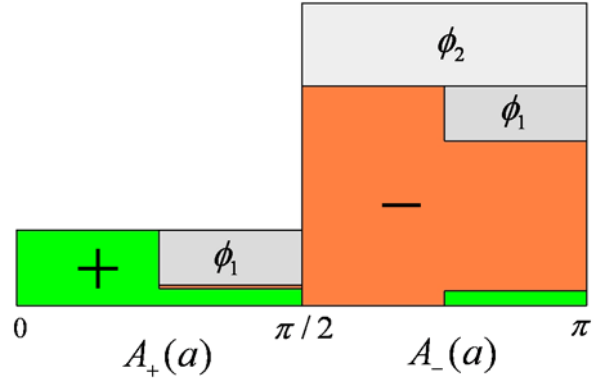

**FIG. 5.** 2x2 LHV model for non-maximally entangled quantum states. The green (+) and orange (-) colored polygons represent subset $B_+(b)$ and $B_-(b)$, and the grey background ($\phi_1$ and $\phi_2$) represents subset $B^c(b)$. The parameters used to construct this figure are $r = 0.26$, $a_i = 0, \pi/4$, $b_j = \pm\pi/16$.

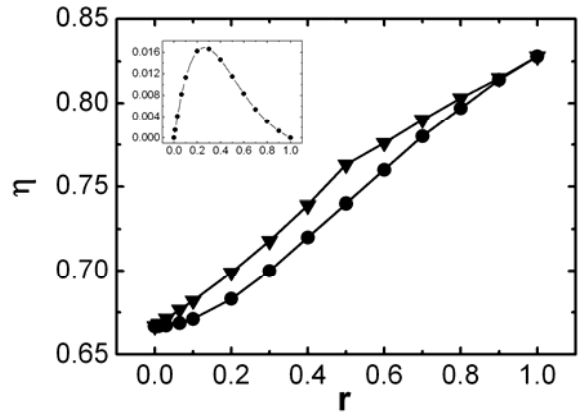

**FIG. 6.** Critical efficiency $\eta^{LHV}(r)$ from the 2x2 LHV model (circles) in Fig 5 versus critical efficiency $\eta^{CH}(r)$ from the CH-Eberhard inequality (triangles). The small figure inside shows the difference between $\eta^{CH}(r)$ and $\eta^{LHV}(r)$.

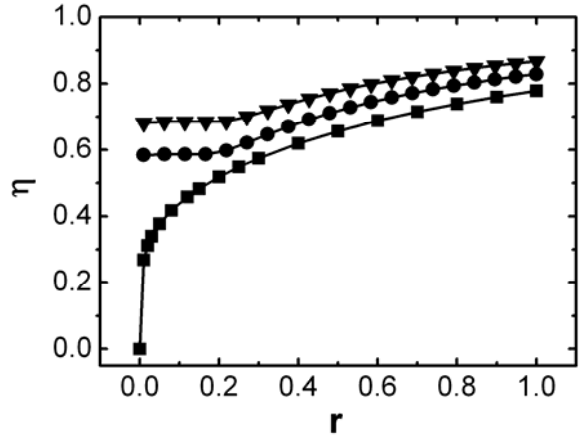

**FIG. 7.** Critical efficiencies $\eta^{LHV}(r)$ from the 2x2 (triangles), the 3x3 (circles), and the NxN (squares) LHV models as functions of the parameter $r$.

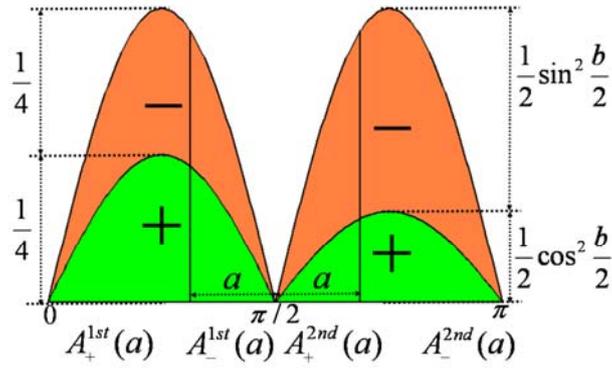

**FIG. 8.** An optimal LHV model for a delayed-choice experiment. The green (+) and orange (-) colored areas represent subset $B_+(b)$ and $B_-(b)$. The parameters used to construct this figure are $a = 5\pi/16$, $b = 5\pi/8$.

27